\documentclass[reprent,aps,prb,twocolumn,superscriptaddress,nolongbibliography]{revtex4-2}

\usepackage{amsmath,amssymb}
\usepackage{graphicx}
\usepackage{bm}
\usepackage{multirow}
\usepackage{hyperref}
\usepackage{wasysym}

\begin{document}

\title{Dynamical generation of skyrmion and bimeron crystals\\by a circularly polarized electric field in frustrated magnets}
\author{Ryota Yambe}
\email{ryota_yambe@alumni.u-tokyo.ac.jp}
\affiliation{Department of Applied Physics, The University of Tokyo, Tokyo 113-8656, Japan }
\author{Satoru Hayami}
\email{hayami@phys.sci.hokudai.ac.jp}
\affiliation{Graduate School of Science, Hokkaido University, Sapporo 060-0810, Japan}

\begin{abstract}
A skyrmion crystal (SkX) has attracted much attention in condensed matter physics, since topologically nontrivial structures induce fascinating physical phenomena.
The SkXs have been experimentally observed in a variety of materials, where the Zeeman coupling to the static magnetic field plays an important role in the formation of the SkXs.
In this study, we theoretically propose another route to generate the SkXs by using a circularly polarized electric field. 
We investigate a non-equilibrium steady state in a classical frustrated Heisenberg magnet under the circularly polarized electric field, where the electric field is coupled to the electric polarization via the spin-current mechanism.  
By numerically solving the Landau-Lifshitz-Gilbert equation at zero temperature, we show that the electric field radiation generates a SkX with a high topological number in the high-frequency regime, where the sign of the skyrmion number is fixed to be negative (positive) under the left (right) circularly polarized field. 
The intense electric field melts these SkXs and generates isolated skyrmions.
We clarify that the microscopic origin is effective electric-field-induced three-spin interactions by adopting the high-frequency expansion in the Floquet formalism. 
Furthermore, we find that the electric field radiation generates another type of SkXs, a bimeron crystal, in the low-frequency regime.
Our results provide a way to generate the SkXs and control the topology by the circularly polarized electric field. 
\end{abstract}

\maketitle

\section{Introduction}
\label{sec:introduction}

Magnetic skyrmions have attracted much attention in condensed matter physics~\cite{nagaosa2013topological,zhang2020skyrmion,gobel2021beyond,hayami2021topological}.
The topologically nontrivial skyrmion structures are characterized by an integer topological number, the so-called skyrmion number $n_{\rm sk}$. 
Their topologically nontrivial structures give rise to fascinating physical phenomena, such as the Hall/Nernst effect~\cite{Ohgushi_PhysRevB.62.R6065,taguchi2001spin,Neubauer_PhysRevLett.102.186602,Shiomi_PhysRevB.88.064409,kurumaji2019skyrmion,Hirschberger_PhysRevLett.125.076602} and the current-induced motion~\cite{Jonietz_skyrmion,yu2012skyrmion,jiang2017direct,yu2020motion}.
Since the observation of the skyrmion with $|n_{\rm sk}|=1$ in MnSi with the chiral cubic symmetry~\cite{Muhlbauer_2009skyrmion}, it was also observed in a variety of materials without the inversion symmetry of the lattice structure~\cite{Tokura_doi:10.1021/acs.chemrev.0c00297}.
In these materials, the skyrmions form a periodic array, the so-called skyrmion crystal (SkX). 
The microscopic origin of the SkX is the synergy of the ferromagnetic exchange interaction, the Dzyaloshinskii-Moriya (DM) interaction~\cite{dzyaloshinsky1958thermodynamic, moriya1960anisotropic}, and the Zeeman coupling to the static magnetic field~\cite{Bogdanov89,Bogdanov94,rossler2006spontaneous}.  
Recently, the SkXs with $|n_{\rm sk}|=1$ were observed under the magnetic filed in materials with the inversion symmetry, such as  Gd$_2$PdSi$_3$~\cite{kurumaji2019skyrmion,Hirschberger_PhysRevLett.125.076602,Hirschberger_PhysRevB.101.220401,Spachmann_PhysRevB.103.184424}, Gd$_3$Ru$_4$Al$_{12}$~\cite{hirschberger2019skyrmion}, GdRu$_2$Si$_2$~\cite{khanh2020nanometric,Yasui2020imaging,khanh2022zoology}, EuAl$_4$~\cite{Shang_PhysRevB.103.L020405,kaneko2021charge,takagi2022square,PhysRevB.106.094421,PhysRevB.107.L020410}, and GdRu$_2$Ge$_2$~\cite{yoshimochi2024multi}. 
Since the DM interaction vanishes under the inversion symmetry, other microscopic origins have been theoretically proposed, such as competing exchange interactions~\cite{Okubo_PhysRevLett.108.017206}, dipolar interactions~\cite{Utesov_PhysRevB.103.064414,utesov2021mean}, and biquadratic spin interactions~\cite{Ozawa_PhysRevLett.118.147205,Hayami_PhysRevB.95.224424} in addition to the Zeeman coupling.

Among the effects of external fields, most studies have focused on the Zeeman coupling to the static magnetic field, while the effects of the laser and the electric field have been recently studied as an alternative possibility of realizing the skyrmions. 
The laser pulse radiation heats the system via the coupling of the electric field and the electrons (lattice vibrations)~\cite{RevModPhys.82.2731}, which generates the isolated skyrmions~\cite{PhysRevLett.110.177205,koshibae2014creation,PhysRevB.95.054421} and the SkXs~\cite{je2018creation,PhysRevLett.120.117201}.
The nonthermal effects of the electric field radiation have also been investigated; the generation of the isolated skyrmion by the inverse Faraday effect~\cite{PhysRevB.99.054423} and the deformation of the SkX by the electric-field-induced DM interaction~\cite{PhysRevLett.119.147202}.  
Since the electric field is directly coupled to the spins in multiferroic materials, the electric field radiation has various effects on the skyrmions; in Cu$_2$OSeO$_3$ with the $d$-$p$ hybridized mechanism~\cite{seki2012observation}, rotation of the SkX by the static electric field~\cite{white2012electric,PhysRevLett.113.107203}, control of the SkX phase stability by the static electric field~\cite{okamura2016transition}, and the generation of the skyrmions  by the electric field pulse~\cite{mochizuki2015writing,mochizuki2016creation,huang2018situ}.

In this paper, we theoretically propose another way to generate the SkXs in multiferroic materials by using a circularly polarized electric field. 
Radiation of the circularly polarized electric field with high frequencies effectively induces three-spin interactions coupled to a scalar spin chirality, when the electric field is coupled to the electric polarization via the spin-current mechanism~\cite{PhysRevB.108.064420,yambe2023scalar}. 
Since the SkXs exhibit nonzero scalar spin chirality, such an effect opens another route to generate the SkXs.    
For that purpose, we study the effect of circularly polarized electric field in a classical Heisenberg model on the triangular lattice.
We consider frustrated competing exchange interactions so that the ground state becomes a single-$Q$ spiral state.
We irradiate the single-$Q$ spiral state with the circularly polarized electric field and calculate a time evolution by numerically solving the Landau-Lifshitz-Gilbert (LLG) equation at zero temperature. 
As a result, we clarify that SkXs with a high topological number $|n_{\rm sk}|=2$ are generated in non-equilibrium steady states (NESSs) in the high-frequency regime.
We show that the microscopic origin is accounted for by the electric-field-induced three-spin interactions by adopting the high-frequency expansion in the Floquet formalism.
By increasing the electric field, the SkXs melt and isolated skyrmions are generated.
We also study the low-frequency regime, where the Floquet analysis in the high-frequency regime is not valid. 
There, we find another type of SkXs with $|n_{\rm sk}|=1$, the so-called bimeron crystal.
Our results show that the two types of SkXs are dynamically generated by the circularly polarized electric field depending on the frequency regime.
Furthermore, we reveal that the sign of the skyrmion number is controlled by the polarization of the electric field in both high- and low-frequency regimes. 

The paper is organized as follows. 
In Sec.~\ref{sec:model}, we introduce a classical Heisenberg model as a static model.
We also introduce a dynamical model including the effect of a circularly polarized electric field.
Then, we outline the method based on the LLG equation in Sec.~\ref{sec:method}.   
We show the simulation results for the left circular polarization (LCP) in the high- and low-frequency regimes in Secs.~\ref{sec:simulation_high} and \ref{sec:simulation_low}, respectively.
Section~\ref{sec:summary} is devoted to the summary and the discussion. 
The simulation results for the right circular polarization (RCP) are shown in Appendix~\ref{sec:RCP}.
The details of the Floquet analysis are given in Appendix~\ref{sec:derivation}.

\section{Model}
\label{sec:model}

We consider a classical Heisenberg model on the two-dimensional triangular lattice, whose static Hamiltonian is given by
\begin{align}
\label{eq:static_hamiltonian}
\mathcal{H}_0&=
-J_1\sum_{\langle j,k \rangle}\boldsymbol{m}_j\cdot\boldsymbol{m}_k
-J_3\sum_{\langle\langle j,k \rangle\rangle}\boldsymbol{m}_j\cdot\boldsymbol{m}_k.
\end{align}
Here, $\boldsymbol{m}_j$ is the local magnetic moment at site $j$ with $|\boldsymbol{m}_j|=1$ and $J_1>0$ ($J_3<0$) represents the ferromagnetic (antiferromagnetic) exchange interaction between the nearest-neighbor pairs $\langle j,k \rangle$ (third-neighbor pairs $\langle\langle j,k \rangle\rangle$). 
In the model with $J_3<-J_1/4$, the ground state becomes a single-$Q$ spiral state, whose ordering wave vector is given by $\boldsymbol{q}^*_1=q^*(1,0)$, $\boldsymbol{q}^*_2=q^*(-1/2,\sqrt{3}/2)$ or $\boldsymbol{q}^*_3=q^*(-1/2,-\sqrt{3}/2)$ with $q^*=2\cos^{-1}\{(1+\sqrt{1-2J_1/J_3})/4\}$; we set the lattice constant as the length unit.
To stabilize the SkXs in the ground state, single-ion anisotropy~\cite{leonov2015multiply, Lin_PhysRevB.93.064430, Hayami_PhysRevB.93.184413, Hayami_PhysRevB.103.224418} or anisotropic exchange interactions~\cite{amoroso2020spontaneous,amoroso2021tuning} are required in addition to the Zeeman coupling to the static magnetic field.  
At finite temperatures, the SkXs are formed by the synergy of the thermal fluctuations and the static magnetic field when $|J_3|$ is larger than $J_1$~\cite{Okubo_PhysRevLett.108.017206}.

We investigate another route for realizing the SkXs by introducing a circularly polarized electric field to the static Hamiltonian in Eq.~(\ref{eq:static_hamiltonian}), whose dynamical Hamiltonian is given by 
\begin{align}
\label{eq:dynamical_hamiltonian}
\mathcal{H}(t) = \mathcal{H}_0 -\bm{E}(t)\cdot\sum_{\langle j,k \rangle}\boldsymbol{p}_{jk},
\end{align}
where the second term represents the coupling of the electric field $\bm{E}(t)$ and the electric dipole $\boldsymbol{p}_{jk}$ on the nearest-neighbor bond.
The circularly polarized electric field is given by $\bm{E}(t) = E_0(\delta\cos\Omega t,-\sin\Omega t ,0)$, where $E_0$ is the amplitude, $\Omega$ is the frequency, and $\delta=+1$ $(-1)$ stands for the RCP (LCP). 
We suppose that the electric dipole is induced by the spin-current mechanism as  $\bm{p}_{jk}=-\lambda\bm{e}_{jk} \times (\bm{m}_j\times \bm{m}_k)$~\cite{Katsura_PhysRevLett.95.057205,Mostovoy_PhysRevLett.96.067601,SergienkoPhysRevB.73.094434}, where $\lambda$ is the magnetoelectric coupling constant and $\bm{e}_{jk}$ is the unit vector from the site $j$ to the site $k$.
In multiferroic materials, the electric dipole by the spin current mechanism has been experimentally observed~\cite{tokura2014multiferroics}. 
It is noted that the form of the dynamical Hamiltonian in Eq.~(\ref{eq:dynamical_hamiltonian}) is the most general form from a viewpoint of symmetry, since all the terms are allowed under the hexagonal point group symmetry $D_{6h}$ of the triangular lattice.

\section{Method}
\label{sec:method}

We study the NESSs in the dynamical model in Eq.~(\ref{eq:dynamical_hamiltonian}) by numerically solving the LLG equation at zero temperature.
The LLG equation is given by
\begin{align}
\label{eq:LLG}
\frac{d\bm{m}_j}{dt}=&-\frac{\gamma}{1+\alpha_{\rm G}^2}\left[ \bm{m}_j\times\bm{B}^{\rm eff}_j(t) \right. \nonumber \\
& \left.+ \alpha_{\rm G}\bm{m}_j\times\{\bm{m}_j\times\bm{B}^{\rm eff}_j(t)\} \right],
\end{align}
where $\gamma$, $\alpha_{\rm G}$, and $\bm{B}^{\rm eff}_{j}(t)=-\partial\mathcal{H}(t)/\partial \bm{m}_{j}$ represent the gyromagnetic ratio, the Gilbert damping constant, and the effective magnetic field, respectively.
The first term with $\gamma$ represents the precession around the effective magnetic field.
Meanwhile, the second term with $\alpha_{\rm G}$ describes the relaxation to the effective magnetic field, which makes the system approach the NESSs.
It is noted that the heating effect~\cite{PhysRevLett.110.177205,koshibae2014creation,PhysRevB.95.054421,je2018creation,PhysRevLett.120.117201} is ignored here.

We calculate magnetization, scalar spin chirality, skyrmion number, and magnetic moment in momentum space as a function of time.
The magnetization is given by 
\begin{align}
M^\alpha(t) = \frac{1}{N}\sum_{j}m^\alpha_j(t),
\end{align}
where $N$ is the system size and $\alpha=x,y,z$. 
We also calculate an in-plane magnetization by $M^{xy}(t)=\sqrt{\{M^x(t)\}^2+\{M^y(t)\}^2}$.
The scalar spin chirality is given by 
\begin{align}
\chi_{\bm{r}}(t) = \bm{m}_j(t)\cdot\{ \bm{m}_k(t)\times \bm{m}_l(t)\}
\end{align}
where the position vector $\bm{r}$ represents the center of the triangle and the sites $(j,k,l)$ are labeled in the counterclockwise order.
The skyrmion number for the whole system is given by
\begin{align}
\label{eq:Nsk}
N_\mathrm{sk}(t) = \frac{1}{4\pi} \sum_{\bm{r}} \Omega_{\bm{r}}(t),
\end{align}
where $\Omega_{\bm{r}}(t)\in [-2\pi,2\pi)$ is a skyrmion density~\cite{BERG1981412};
\begin{align}
\tan\frac{\Omega_{\bm{r}}(t)}{2} = \frac{2\bm{m}_j(t)\cdot\{ \bm{m}_k(t)\times \bm{m}_l(t)\}}{\{\bm{m}_j(t)+\bm{m}_k(t)+\bm{m}_l(t)\}^2-1}.
\end{align}
We also calculate the skyrmion number for the magnetic unit cell as $n_{\rm sk}(t)=N_{\rm sk}(t)/N_{\rm MU}$ with the number of magnetic unit cells $N_{\rm MU}$, which characterizes the topological property of the constituent skyrmions.
The magnetic moment with the wave vector $\bm{q}$ is defined as 
\begin{align}
m_{\bm{q}}(t) = \sqrt{\frac{1}{N^2}\sum_{\alpha,j,k}m^\alpha_j(t)m^\alpha_k(t) e^{i\bm{q}\cdot(\bm{R}_j-\bm{R}_k)}},
\end{align}
where $\bm{R}_j$ is the position vector of the site $j$.
After the system reaches the NESS at a long time $t_0$, we average the above physical quantities $O(t)$ over time as
\begin{align}
\label{eq:average}
O &= \frac{1}{N_{\rm ave}}\sum_{n=1}^{N_{\rm ave}} O(t_0+n\Delta).
\end{align} 
Here, $N_{\rm ave}$ and $\Delta$ represent the number of samples and the time step, respectively.

\section{Simulation results}
\label{sec:simulation}

We investigate the effect of the electric field radiation on the single-$Q$ spiral state by numerically solving the LLG equation.
We set $J_1=1$ and $J_3=-0.5$ in the static Hamiltonian in Eq.~(\ref{eq:static_hamiltonian}); the ground state becomes the single-$Q$ spiral state with $q^*=2\pi/5$ and its energy per site is given by $E_{1Q}=-J_1\{ \cos q^* +2\cos (q^*/2) \}-J_3\{ \cos (2q^*) +2\cos q^* \}$. 
It is noted that there is no instability toward the SkXs within the static Hamiltonian even when the static magnetic field and the thermal fluctuations are additionally applied.
The energy scale and the time scale of the system are characterized by $|E_{1Q}|$ and $|E_{1Q}|^{-1}$, respectively.
We irradiate the single-$Q$ spiral state with the circularly polarized electric field and calculate a time evolution for a long time $t_0=4000T$, where $T=2\pi/\Omega$ is a time period of the electric field.
Then, we calculate the averaged physical quantities according to the definition in Eq.~(\ref{eq:average}), where we set $N_{\rm ave}=400$ and $\Delta=0.02T$.
We solve the LLG equation with $\gamma=1$, $\alpha_{\rm G}=0.05$, and $N=20^2$~\footnote[1]{We confirm that similar results are obtained for larger system sizes, e.g., $N=50^2$.} under the periodic boundary condition by using an open software DifferentialEquations.jl~\cite{rackauckas2017differentialequations}. 
In the following, we show the results in the high-frequency (low-frequency) regime in Sec.~\ref{sec:simulation_high} (\ref{sec:simulation_low}).

\subsection{High frequency regime}
\label{sec:simulation_high}

\begin{figure}[t!]
\begin{center}
\includegraphics[width=1.0\hsize]{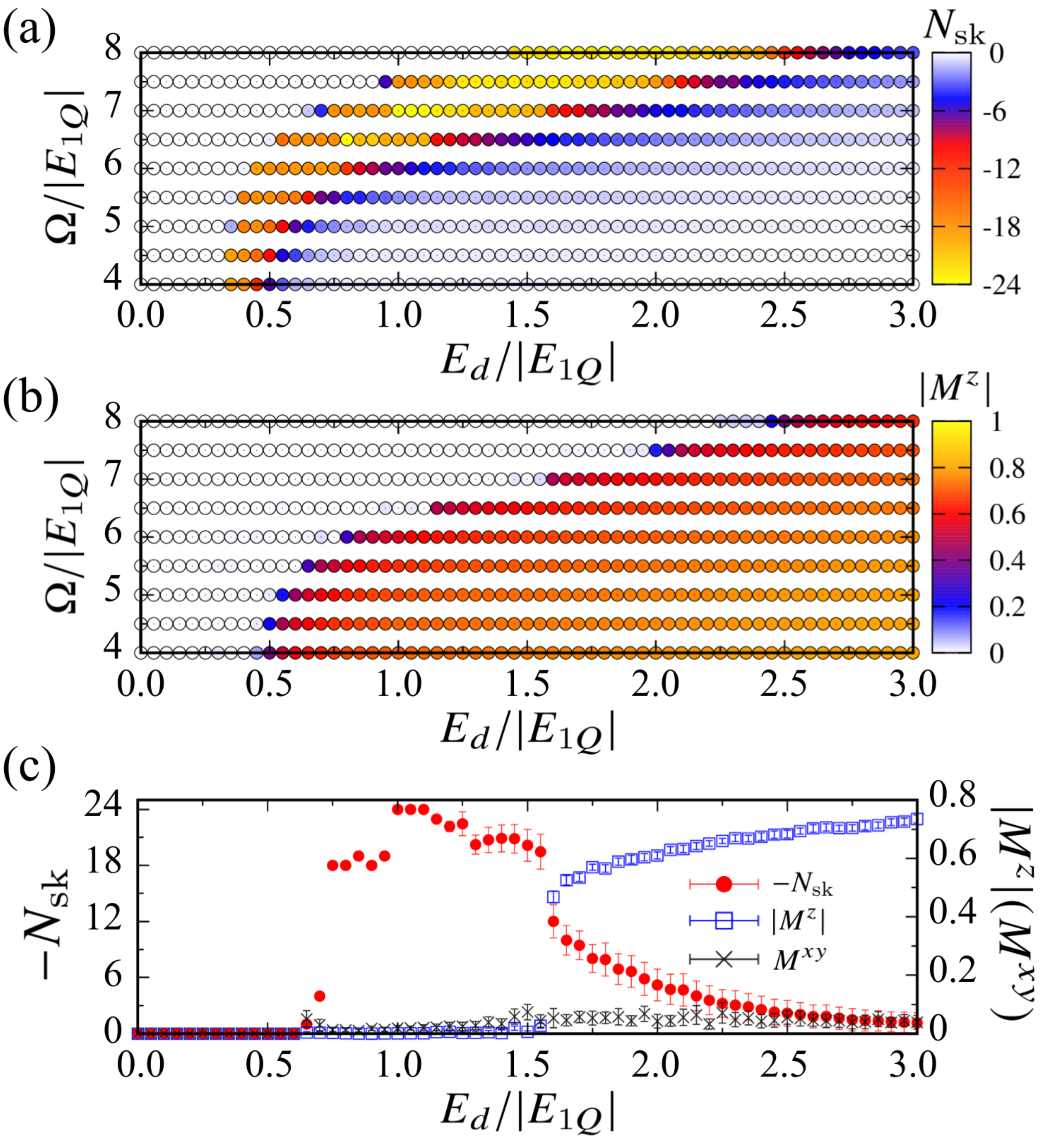} 
\caption{\label{fig:high}
(a) Averaged skyrmion number $N_{\rm sk}$ and (b) averaged out-of-plane magnetization $M^z$ as functions of $E_{d}$ and $\Omega$ in the NESSs by the left circularly polarized electric fields in the high-frequency regime.
(c) $E_{d}$ dependence of $N_{\rm sk}$, $M^z$, and the averaged in-plane magnetization $M^{xy}$ at $\Omega/|E_{1Q}|=7$. 
}
\end{center}
\end{figure}

First, we study the high-frequency regime, where the frequency $\Omega$ is large enough compared to the static energy scale $|E_{1Q}|$.
The initial state is set as a single-$Q$ horizontal spiral state with the ordering wave vector $\boldsymbol{q}^* _1$, whose spiral plane lies on the $xy$ plane~\footnote[2]{Similar results are obtained when the initial state is set as the other single-$Q$ spiral states with the different ordering wave vectors and spiral planes.}.
We show the averaged skyrmion number $N_{\rm sk}$ and the averaged out-of-plane magnetization $M^z$ with changing $E_d=\lambda E_0$ and $\Omega$ in Figs.~\ref{fig:high}(a) and \ref{fig:high}(b), respectively, where we radiate the electric field with the LCP.  
The details of the $E_d$ dependence at $\Omega/|E_{1Q}|=7$ are shown in Fig.~\ref{fig:high}(c).

At $E_d=0$ in Fig.~\ref{fig:high}(c), the skyrmion number, the out-of-plane magnetization, and the in-plane magnetization are zero due to the appearance of the single-$Q$ spiral structure.
They remain zero even when the weak electric field ($E_d/|E_{1Q}|\le 0.6$) is applied.
Meanwhile, the relatively large electric field ($E_d/|E_{1Q}|\ge 0.65 $) modulates the single-$Q$ spiral structure and induces $N_{\rm sk}$,  $M^z$, and $M^{xy}$, as shown in Fig.~\ref{fig:high}(c).
Here, the sign of $N_{\rm sk}$ is fixed to be negative, but the sign of the magnetization is not fixed.
Similar results are obtained for different $\Omega$, as shown in Figs.~\ref{fig:high}(a) and \ref{fig:high}(b).

\begin{figure*}[t!]
\begin{center}
\includegraphics[width=1.0\hsize]{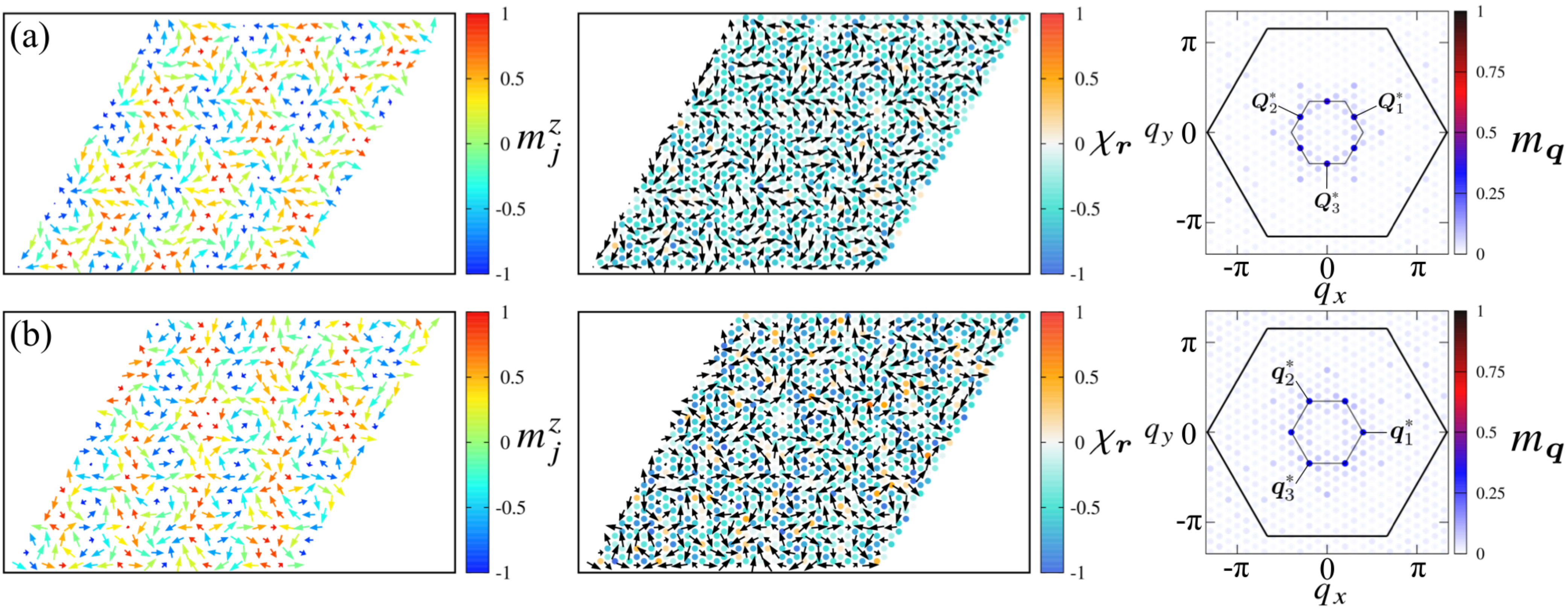} 
\caption{\label{fig:high_conf}
Single snapshots of the NESSs  by the left circularly polarized electric fields with (a) $\Omega/|E_{1Q}|=7$ and $E_d/|E_{1Q}|=0.75$ and (b) $\Omega/|E_{1Q}|=7$ and $E_d/|E_{1Q}|=1.0$.
(Left panels) Magnetic configurations; the arrows and color denote the magnetic moments $\bm{m}_j(t)$ and the $z$ component $m^z_j(t)$, respectively.
(Middle panels) Scalar spin chirality configurations; the arrows and color of the circles denote the magnetic moments $\bm{m}_j(t)$ and the scalar spin chirality $\chi_{\bm{r}}(t)$, respectively.
(Right panels) Magnetic moments in momentum space $m_{\bm{q}}(t)$.
The large hexagons represent the first Brillouin zone.
The vertices of the small hexagons correspond to $\pm\boldsymbol{q}^*_1$, $\pm\boldsymbol{q}^*_2$, and $\pm\boldsymbol{q}^*_3$.
}
\end{center}
\end{figure*}

For $0.65 \le E_d/|E_{1Q}| \le 1.1$ in Fig.~\ref{fig:high}(c), we find that SkXs are generated in the NESSs with $N_{\rm sk}=-18$ and $-24$.
We show the magnetic configuration of the SkX with $N_{\rm sk}=-18$ at $E_d/|E_{1Q}|=0.75$ in the left panel of Fig.~\ref{fig:high_conf}(a).
The SkX consists of vortices with negative $m^z_j$ and antivortices with positive $m^z_j$.
Both vortices and antivortices exhibit the negative scalar spin chirality, as shown in the 
middle panel of Fig.~\ref{fig:high_conf}(a). 
The periodic structure is characterized by the three ordering wave vectors $\boldsymbol{Q}^*_1=(\boldsymbol{q}^*_1+\boldsymbol{q}^*_2)/2$, $\boldsymbol{Q}^*_2=(\boldsymbol{q}^*_2+\boldsymbol{q}^*_3)/2$, and $\boldsymbol{Q}^*_3=(\boldsymbol{q}^*_3+\boldsymbol{q}^*_1)/2$, as shown in the right panel of Fig.~\ref{fig:high_conf}(a). 
The magnetic unit cell includes the two vortices and the two antivortices; its magnetic configuration corresponds to a skyrmion with $n_{\rm sk}=-2$, the so-called higher-order skyrmion.
The periodic array of the higher-order skyrmions, the higher-order SkX, includes the nine higher-order skyrmions in the whole system, resulting in $N_{\rm sk}=-18$. 
The SkXs with $N_{\rm sk}=-24$ are generated in the larger field region than the SkXs with $N_{\rm sk}=-18$.
Similar to the SkX with $N_{\rm sk}=-18$, this state is the higher-order SkX, as shown in Fig.~\ref{fig:high_conf}(b).
Meanwhile, the ordering wave vectors are given by $\boldsymbol{q}^*_1$, $\boldsymbol{q}^*_2$, and $\boldsymbol{q}^*_3$ rather than $\boldsymbol{Q}^*_1$, $\boldsymbol{Q}^*_2$, and $\boldsymbol{Q}^*_3$.
The magnetic unit cell becomes smaller compared to the SkXs with $N_{\rm sk}=-18$, resulting in the twelve higher-order skyrmions in the whole system.
It is noted that similar higher-order SkXs are stabilized in static models by biquadratic spin interactions~\cite{Ozawa_PhysRevLett.118.147205,Hayami_PhysRevB.95.224424} and symmetric off-diagonal anisotropic exchange interactions~\cite{amoroso2020spontaneous,amoroso2021tuning,yambe2021skyrmion}, although the present mechanism requires neither such multiple-spin and anisotropic interactions.

The higher-order SkXs are generated by the electric field with different $\Omega$, as shown in Fig.~\ref{fig:high}(a).
The phase diagram shows that the larger $E_d$ is required for larger $\Omega$ in order to generate the SkXs. 
The higher-order SkXs are also generated by the electric field with the RCP, where the sign of $N_{\rm sk}$ is opposite to the case with the LCP, as shown in Appendix~\ref{sec:RCP}; the electric field with the LCP (RCP) generates the higher-order SkXs with negative (positive) $N_{\rm sk}$.

Let us discuss the microscopic origin of the higher-order SkX based on the Floquet formalism~\cite{eckardt2017colloquium,oka2019floquet,rudner2020floquet}.
In the Floquet formalism, the effect of a time-periodic field is analyzed by using a time-independent Floquet Hamiltonian.
The Floquet Hamiltonian for classical spin systems is obtained by expanding the LLG equation under the time-periodic field with respect to $\Omega^{-1}$ when the frequency is much larger than the static energy scale~\cite{higashikawa2018floquet}. 
Following this method, we derive the Floquet Hamiltonian, which is given by
\begin{align}
\label{eq:floquet_hamiltonian}
\mathcal{H}_F &= \mathcal{H}_0 + \mathcal{T}\sum_{\bigtriangleup,\bigtriangledown}\boldsymbol{m}_j\cdot\{\boldsymbol{m}_k\times\boldsymbol{m}_l\} \nonumber\\
& + \mathcal{T}\sum_{j}\sum_{\sigma=\pm1}m_j^z\{\boldsymbol{m}_{j+\sigma\bm{e}_1}\times\boldsymbol{m}_{j+\sigma\bm{e}_2} \nonumber\\
&+ \boldsymbol{m}_{j+\sigma\bm{e}_2}\times\boldsymbol{m}_{j+\sigma\bm{e}_3} + \boldsymbol{m}_{j+\sigma\bm{e}_3}\times\boldsymbol{m}_{j+\sigma\bm{e}_1} \}^z,
\end{align}
with 
\begin{align}
\label{eq:T}
\mathcal{T} = -\frac{\sqrt{3}\delta\gamma E_d^2}{4\Omega(1+\alpha_{\rm G}^2)}.
\end{align}
The detailed derivation is presented in Appendix~\ref{sec:derivation}.
The first term in Eq.~(\ref{eq:floquet_hamiltonian}) corresponds to Eq.~(\ref{eq:static_hamiltonian}) and the other terms are effective electric-field-induced three-spin interactions; the summation in the second term is taken over the upward and downward unit triangles and the sites $(j,k,l)$ are labeled in the counterclockwise order; in the third term, $\bm{e}_1=(1,0)$, $\bm{e}_2=(-1/2,\sqrt{3}/2)$, and $\bm{e}_3=(-1/2,-\sqrt{3}/2)$ represent the vectors of the nearest-neighbor bonds.
The three-spin interactions are not induced when the electric field is linearly polarized ($\delta=0$).
This is understood from the fact that the time-reversal symmetry breaking by the circular polarization is necessary for inducing the three-spin interactions with the time-reversal odd symmetry. 
Similar three-spin interactions were studied in Mott insulators~\cite{PhysRevB.96.014406,claassen2017dynamical,quito2021polarization,PhysRevB.105.054423},  quantum spin systems~\cite{sato2014floquet,PhysRevB.108.064420}, and a classical kagome magnet~\cite{yambe2023scalar}.

The emergence of the SkXs in Fig.~\ref{fig:high}(a) is qualitatively understood from the effective field-induced three-spin interactions in Eqs.~(\ref{eq:floquet_hamiltonian}) and (\ref{eq:T}).
From the expression in Eq.~(\ref{eq:T}), one finds that the radiation of the electric field with the LCP ($\delta=-1$) effectively induces the three spin interactions with $\mathcal{T}>0$, which favors the negative scalar spin chirality and skyrmion number.
This is in agreement with the simulation results in Figs.~\ref{fig:high} and \ref{fig:high_conf}.
According to Eq.~(\ref{eq:T}), the amplitude of $\mathcal{T}$ is proportional to $E_d^2$.
This is why $N_{\rm sk}$ increases by increasing $E_d$.
In addition, the amplitude of $\mathcal{T}$ is smaller when the frequency $\Omega$ is larger.
This accounts for the behavior that larger $E_d$ is required to induce $N_{\rm sk}$ for larger $\Omega$.   
When the polarization is reversed so as to have the RCP ($\delta=+1$), the sign of $\mathcal{T}$ becomes negative; the negative three-spin interactions favor the positive scalar spin chirality and skyrmion number, which is also consistent with the results shown in Appendix~\ref{sec:RCP}. 
It is noted that analysis based on the Floquet Hamiltonian in Eq.~(\ref{eq:floquet_hamiltonian}) is no longer valid by considering the larger $E_d$ region, where the SkX melts, as will be discussed below, since the higher-order terms proportional to $E_d^2\Omega^{-n}$ ($n \ge 2$) are non-negligible.

\begin{figure}[t!]
\begin{center}
\includegraphics[width=1.0\hsize]{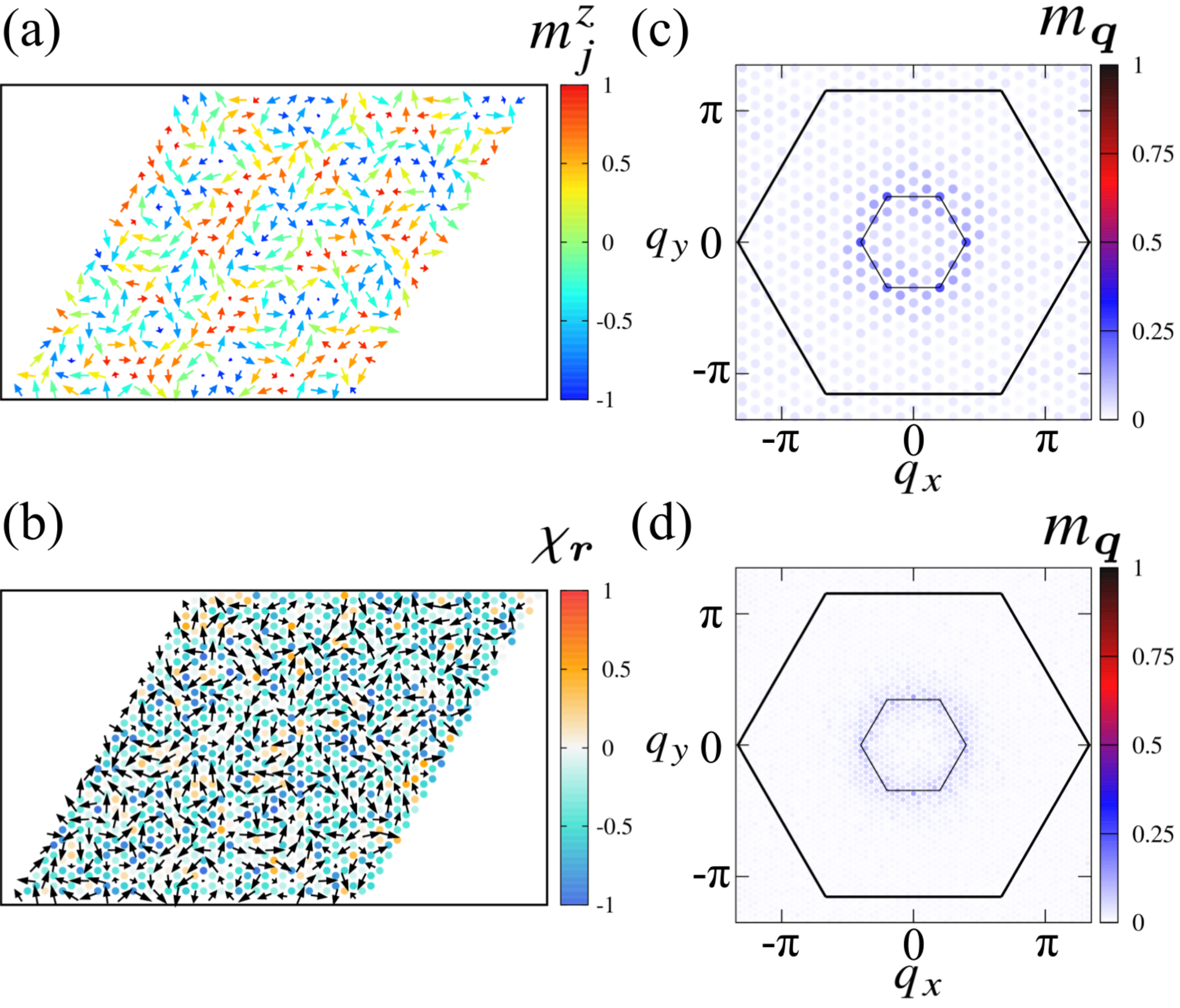} 
\caption{\label{fig:high_melt}
Single snapshots of the NESSs  by the left circularly polarized electric fields with $\Omega/|E_{1Q}|=7$ and $E_d/|E_{1Q}|=1.15$.
(a) Magnetic configurations; the arrows and color denote the magnetic moments $\bm{m}_j(t)$ and the $z$ component $m^z_j(t)$, respectively.
(b) Scalar spin chirality configurations; the arrows and color of the circles denote the magnetic moments $\bm{m}_j(t)$ and the scalar spin chirality $\chi_{\bm{r}}(t)$, respectively.
(c) [(d)] Magnetic moments in momentum space $m_{\bm{q}}(t)$ for $N=20^2$ ($N=50^2$).
The large hexagons represent the first Brillouin zone.
The vertices of the small hexagons correspond to $\pm\boldsymbol{q}^*_1$, $\pm\boldsymbol{q}^*_2$, and $\pm\boldsymbol{q}^*_3$.
}
\end{center}
\end{figure}

The averaged $N_{\rm sk}$ shows a slight decrease for $1.15\le E_d/|E_{1Q}| \le 1.55$ in Fig.~\ref{fig:high}(c) while the small magnetization is kept.
The snapshots of the NESS at $E_d/|E_{1Q}|=1.15$ are shown in Fig.~\ref{fig:high_melt}.
Similar to the higher-order SkXs in Fig.~\ref{fig:high_conf}, the vortices and antivorticies with the negative scalar spin chirality appear in the NESS, as shown in  Figs.~\ref{fig:high_melt}(a) and \ref{fig:high_melt}(b).
Meanwhile, the magnetic moments in momentum space in Fig.~\ref{fig:high_melt}(c) show the broad distribution around the small hexagon rather than the six sharp peaks for the higher-order SkXs.
The broad distribution becomes clearer by increasing the system size, as shown in Fig.~\ref{fig:high_melt}(d).
This indicates the emergence of skyrmion liquids or spiral spin liquids~\cite{Gao2016Spiral,PhysRevB.99.064435,huang2020melting,PhysRevB.106.224406,PhysRevLett.130.106703,PhysRevB.109.064426} as a result of the melting of the higher-order SkXs by the large electric field, although a further investigation for larger system sizes is required to conclude whether such liquid states remain stable.

\begin{figure}[t!]
\begin{center}
\includegraphics[width=1.0\hsize]{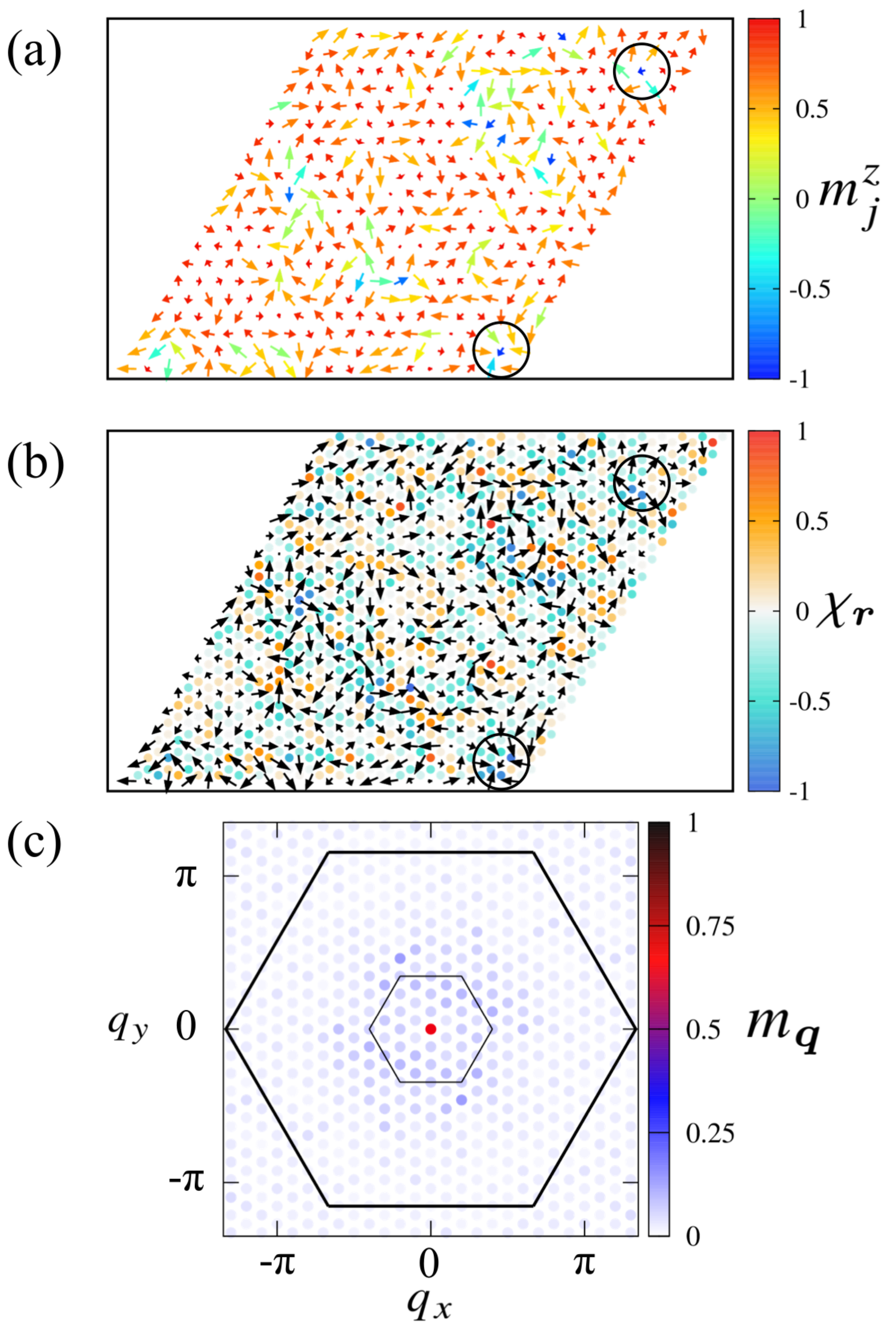} 
\caption{\label{fig:single}
Single snapshots of the NESSs by the left circularly polarized electric fields with $\Omega/|E_{1Q}|=7$ and $E_d/|E_{1Q}|=2.0$.
(a) Magnetic configurations; the arrows and color denote the magnetic moments $\bm{m}_j(t)$ and the $z$ component $m^z_j(t)$, respectively.
(b) Scalar spin chirality configurations; the arrows and color of the circles denote the magnetic moments $\bm{m}_j(t)$ and the scalar spin chirality $\chi_{\bm{r}}(t)$, respectively.
In (a) and (b), the black circles highlight the isolated skyrmions.
(c) Magnetic moments in momentum space $m_{\bm{q}}(t)$.
The large hexagons represent the first Brillouin zone.
The vertices of the small hexagons correspond to $\pm\boldsymbol{q}^*_1$, $\pm\boldsymbol{q}^*_2$, and $\pm\boldsymbol{q}^*_3$.
}
\end{center}
\end{figure}

By further increasing the electric field, the out-of-plane magnetization grows dramatically at $E_d/|E_{1Q}| = 1.6$; for $E_d/|E_{1Q}| \ge 1.6$, $M^z$ ($N_{\rm sk}$) increases (decreases) with increasing $E_d$.
We show the snapshots of the NESS at $E_d/|E_{1Q}|=2.0$ in Fig.~\ref{fig:single}.
The magnetic configuration is mainly characterized by the out-of-plane ferromagnetic one, as shown in Fig.~\ref{fig:single}(a).
We find the isolated skyrmions in the ferromagnetic background, which is highlighted by the black circles.
The isolated skyrmions exhibit the negative scalar chirality, as shown in Fig.~\ref{fig:single}(b).
Due to the ferromagnetic structure, the magnetic moments in momentum space have a peak at $\boldsymbol{q}=\bm{0}$, as shown in Fig.~\ref{fig:single}(c).

\subsection{Low frequency regime}
\label{sec:simulation_low}

\begin{figure}[t!]
\begin{center}
\includegraphics[width=1.0\hsize]{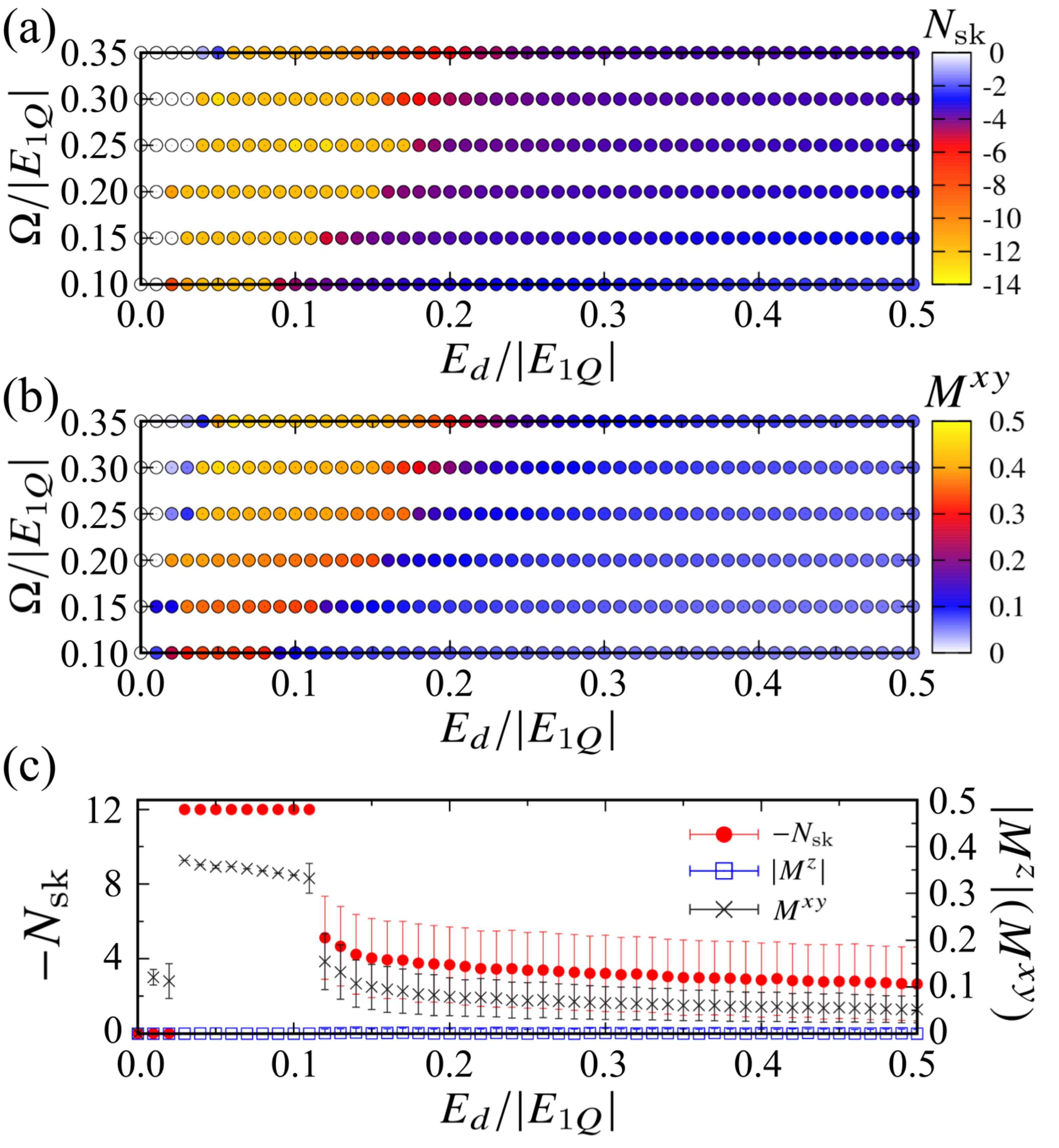} 
\caption{\label{fig:low}
(a) Averaged skyrmion number $N_{\rm sk}$ and (b) averaged in-plane magnetization $M^{xy}$ as functions of $E_{d}$ and $\Omega$ in the NESSs by the left circularly polarized electric fields in the low-frequency regime.
(c) $E_{d}$ dependence of $N_{\rm sk}$, $M^{xy}$, and the averaged out-of-plane magnetization $M^{z}$ at $\Omega/|E_{1Q}|=0.15$.
}
\end{center}
\end{figure}

Next, we study the low-frequency regime, where the frequency $\Omega$ is small enough compared to the static energy scale $|E_{1Q}|$.
We irradiate the single-$Q$ horizontal spiral state with the ordering wave vector $\boldsymbol{q}^* _1$ by the electric field with the LCP~\footnotemark[2].
The $E_d$ and $\Omega$ dependences of the averaged skyrmion number $N_{\rm sk}$ and the averaged in-plane magnetization $M^{xy}$ are shown in Figs.~\ref{fig:low}(a) and \ref{fig:low}(b), respectively.
The details at $\Omega/|E_{1Q}|=0.15$ are shown in Fig.~\ref{fig:low}(c).

In Fig.~\ref{fig:low}(c), the single-$Q$ spiral state with $N_{\rm sk}=M^z=M^{xy}=0$ is stabilized at $E_d=0$.
By applying the electric field, the skyrmion number and the magnetization are induced.
Similar to the high-frequency regime, the sign of $N_{\rm sk}$ is fixed to be negative, while the sign of the magnetization is not fixed.
Different from the high-frequency regime, the in-plane component of the magnetization is mainly induced in the low-frequency regime.
Similar results are obtained for different $\Omega$, as shown in Figs.~\ref{fig:low}(a) and \ref{fig:low}(b).

\begin{figure}[t!]
\begin{center}
\includegraphics[width=1.0\hsize]{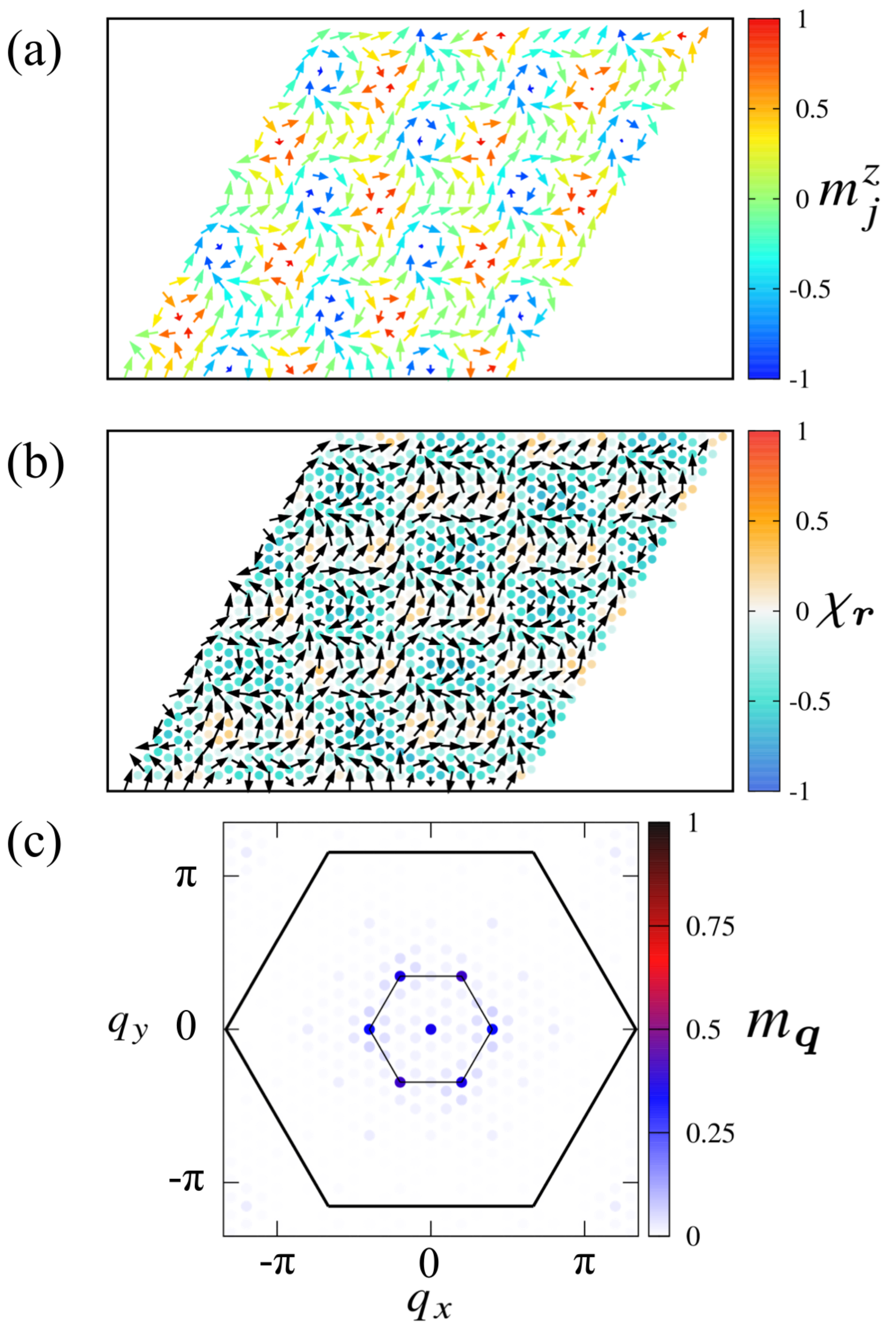} 
\caption{\label{fig:low_conf}
Single snapshots of the NESSs  by the left circularly polarized electric fields with $\Omega/|E_{1Q}|=0.15$ and $E_d/|E_{1Q}|=0.05$.
(a) Magnetic configurations; the arrows and color denote the magnetic moments $\bm{m}_j(t)$ and the $z$ component $m^z_j(t)$, respectively.
(b) Scalar spin chirality configurations; the arrows and color of the circles denote the magnetic moments $\bm{m}_j(t)$ and the scalar spin chirality $\chi_{\bm{r}}(t)$, respectively.
(c) Magnetic moments in momentum space $m_{\bm{q}}(t)$.
The large hexagons represent the first Brillouin zone.
The vertices of the small hexagons correspond to $\pm\boldsymbol{q}^*_1$, $\pm\boldsymbol{q}^*_2$, and $\pm\boldsymbol{q}^*_3$.
}
\end{center}
\end{figure}

The electric field generates SkXs in the NESSs with $N_{\rm sk}=-12$ for $0.03\le E_d/|E_{1Q}| \le 0.11$ at $\Omega/|E_{1Q}|=0.15$, as shown in Fig.~\ref{fig:low}(c).
We show the magnetic configuration at $E_d/|E_{1Q}|=0.05$ in Fig.~\ref{fig:low_conf}(a), which consists of vortices with negative $m^z_j$ and antivortices with positive $m^z_j$.
The vortex and antivortex exhibit the negative scalar spin chirality, as shown in Fig.~\ref{fig:low_conf}(b).   
The ordering wave vectors are given by $\boldsymbol{q}^*_1$, $\boldsymbol{q}^*_2$, and $\boldsymbol{q}^*_3$, as shown in Fig.~\ref{fig:low_conf}(c); the magnetic unit cell consists of a pair of vortex and antivortex, which is the so-called bimeron with $n_{\rm sk}=-1$~\cite{Borge_PhysRevB.99.060407}.
The periodic structure of the bimerons, the bimeron crystal, includes the twelve bimerons in the whole system.
The bimeron crystal shows the large in-plane magnetization, as shown in Fig.~\ref{fig:low}(c), which is also characteristic of the bimeron crystal.
The bimeron crystals are also generated by the electric field with different $\Omega$, as shown in Figs.~\ref{fig:low}(a) and \ref{fig:low}(b).
In addition, we confirm that the bimeron crystals with $N_{\rm sk}=+12$ are generated by the electric field with the RCP, as shown in Appendix~\ref{sec:RCP}.

\begin{figure}[t!]
\begin{center}
\includegraphics[width=1.0\hsize]{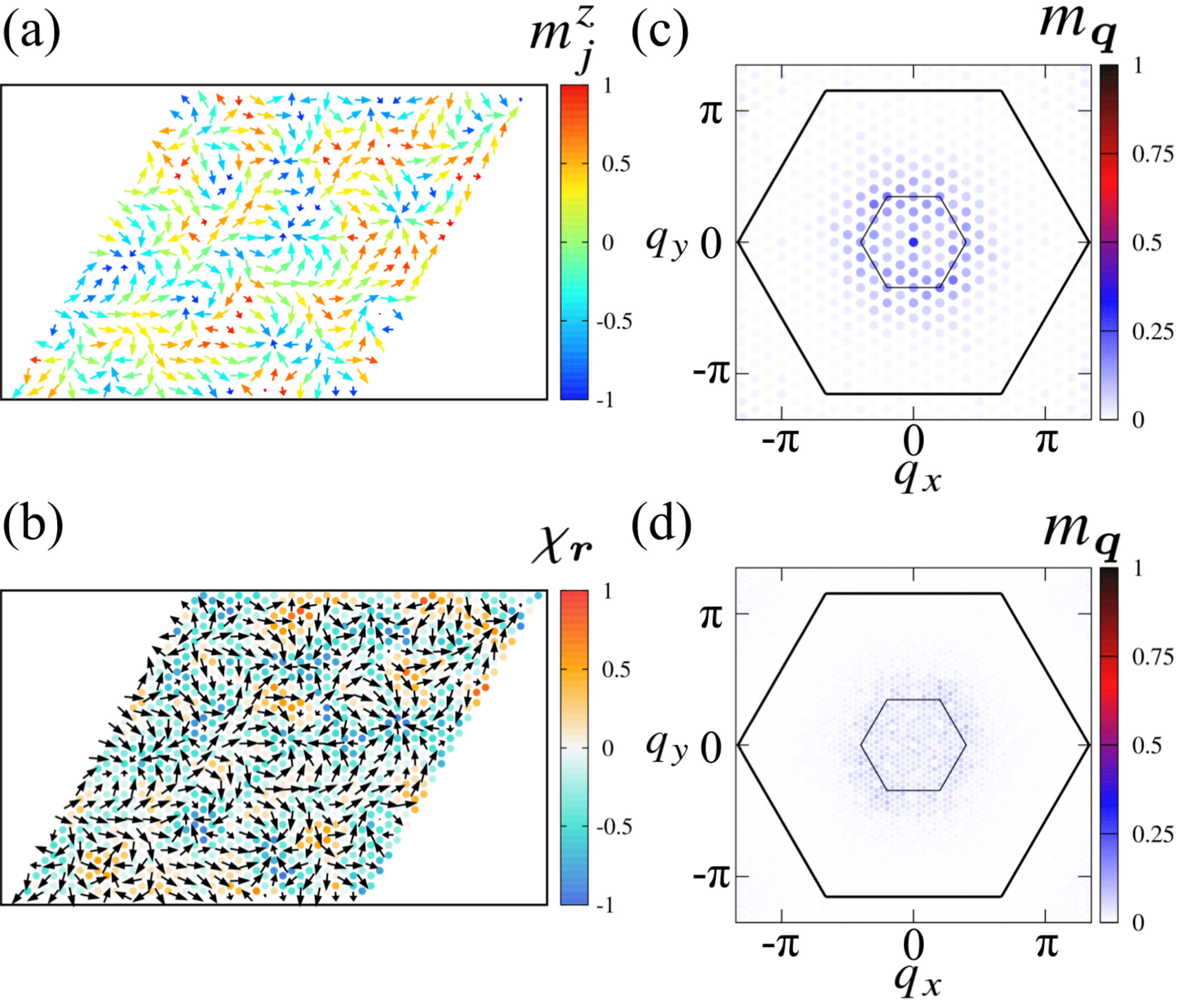} 
\caption{\label{fig:low_melt}
Single snapshots of the NESSs by the left circularly polarized electric fields with $\Omega/|E_{1Q}|=0.15$ and $E_d/|E_{1Q}|=0.13$.
(a) Magnetic configurations; the arrows and color denote the magnetic moments $\bm{m}_j(t)$ and the $z$ component $m^z_j(t)$, respectively.
(b) Scalar spin chirality configurations; the arrows and color of the circles denote the magnetic moments $\bm{m}_j(t)$ and the scalar spin chirality $\chi_{\bm{r}}(t)$, respectively.
(c) [(d)] Magnetic moments in momentum space $m_{\bm{q}}(t)$ for $N=20^2$ ($N=50^2$).
The large hexagons represent the first Brillouin zone.
The vertices of the small hexagons correspond to $\pm\boldsymbol{q}^*_1$, $\pm\boldsymbol{q}^*_2$, and $\pm\boldsymbol{q}^*_3$.
}
\end{center}
\end{figure} 

The averaged $N_{\rm sk}$ and $M^{xy}$ decrease dramatically at $E_d/|E_{1Q}|=0.12$ in Fig.~\ref{fig:low}(c). 
We show the snapshots of the NESS at $E_d/|E_{1Q}|=0.13$ in Fig.~\ref{fig:low_melt}.
Similar to the bimeron crystals, the negative scalar spin chirality is induced in the vortex and antivortex structures, as shown in Figs.~\ref{fig:low_melt}(a) and \ref{fig:low_melt}(b).
Meanwhile, the magnetic moments in momentum space are broadly distributed, as shown in Fig.~\ref{fig:low_melt}(c).
The broad distribution is clearly found by increasing the system size, as shown in Fig.~\ref{fig:low_melt}(d).
This indicates that the bimeron crystals are melted by the large electric field while keeping the local vortex and antivortex structures.

We discuss the microscopic origin of the bimeron crystals based on the previous studies in static Hamiltonians~\cite{Borge_PhysRevB.99.060407,Hayami_PhysRevB.103.224418}, since the Floquet Hamiltonian in Eq.~(\ref{eq:floquet_hamiltonian}) is not valid in the low-frequency regime. 
It has been clarified that the bimeron crystals are stabilized in the ground state by the interplay among the competing exchange interactions, the easy-plane single-ion anisotropy, and the in-plane magnetic field, where the sign of $N_{\rm sk}$ is not fixed.
Since the easy-plane single-ion anisotropy and the in-plane magnetic field are not included in the static Hamiltonian in Eq.~(\ref{eq:static_hamiltonian}), the present mechanism based on the electric field radiation provides a qualitatively different mechanism of the bimeron crystal.
In contrast to the previous mechanism, the electric field radiation fixes the sign of $N_{\rm sk}$ by the polarization of the electric field, as in the high-frequency regime.

\section{Summary and Discussion}
\label{sec:summary}

In summary, we have studied a way to dynamically generate the SkXs in multiferroic materials by using the circularly polarized electric field.
We have irradiated the single-$Q$ spiral state in the classical Heisenberg magnet on the triangular lattice with the circularly polarized electric field, where the electric field is coupled to the electric polarization via the spin-current mechanism. 
We have elucidated that the electric field radiation generates the higher-order SkX and the bimeron crystal with nonzero skyrmion numbers in the high- and low-frequency regimes, respectively, even in the absence of the magnetic field and the heating effect.
The sign of the skyrmion number is fixed to be negative (positive) when the electric field is left (right) circularly polarized.
These SkXs are melted by the intense electric field; the isolated skyrmions are generated especially in the high-frequency regime.
We have also shown that the microscopic origin of the higher-order SkXs in the high-frequency regime is attributed to the electric-field-induced three-spin interactions by adopting the Floquet formalism.
Our results indicate the possibility of generating the SkXs and controlling their topology by the circularly polarized electric field.  

Finally, we discuss a possible experimental situation.
We have studied the classical spin model on the triangular lattice with multiferroic materials in mind, where the single-$Q$ ground state and the spin-current mechanism are key ingredients for the SkX generation.
Such a situation might be realized in a single layer NiI$_2$~\cite{song2022evidence}.  
Assuming an exchange interaction (static energy scale $|E_{1Q}|$) of 1~meV, the frequency $\Omega$ in Figs.~\ref{fig:high} and \ref{fig:low} is in the gigahertz to terahertz regime, where a typical magnitude of the electric field $E_0$ is $1$--$10$ MV/cm.  
The coupling of the electric field and the electric polarization $E_{\rm d}=\lambda E_0$ is estimated to be $10^{-5}E_0$--$10^{-3}E_0$ meV;  $\lambda$ is approximated by $10^{-28}$--$10^{-26}$ $\mu$Cm,  assuming the electric polarization $|\bm{P}|=1$--100 $\mu$C/m$^2$ in multiferroic materials~\cite{tokura2014multiferroics} with the volume of the unit cell 100 \AA$^3$. 
This estimate indicates that our results in the low-frequency regime would be experimentally achievable for multiferroic materials such as NiI$_2$.

\begin{acknowledgments}
We thank S. Okumura, Y. Kato, and Y. Motome for fruitful discussions.
This research was supported by JSPS KAKENHI Grants Numbers JP19K03752, JP19H01834, JP21H01037, JP22H04468, JP22H00101, JP22H01183, JP23KJ0557, JP23H04869, JP23K03288, and by JST PRESTO (JPMJPR20L8) and JST CREST (JPMJCR23O4). 
R.Y. was supported by JSPS Research Fellowship.
\end{acknowledgments}

\appendix
\section{SkXs by the right circularly polarized electric field}
\label{sec:RCP}

\begin{figure*}[t!]
\begin{center}
\includegraphics[width=1.0\hsize]{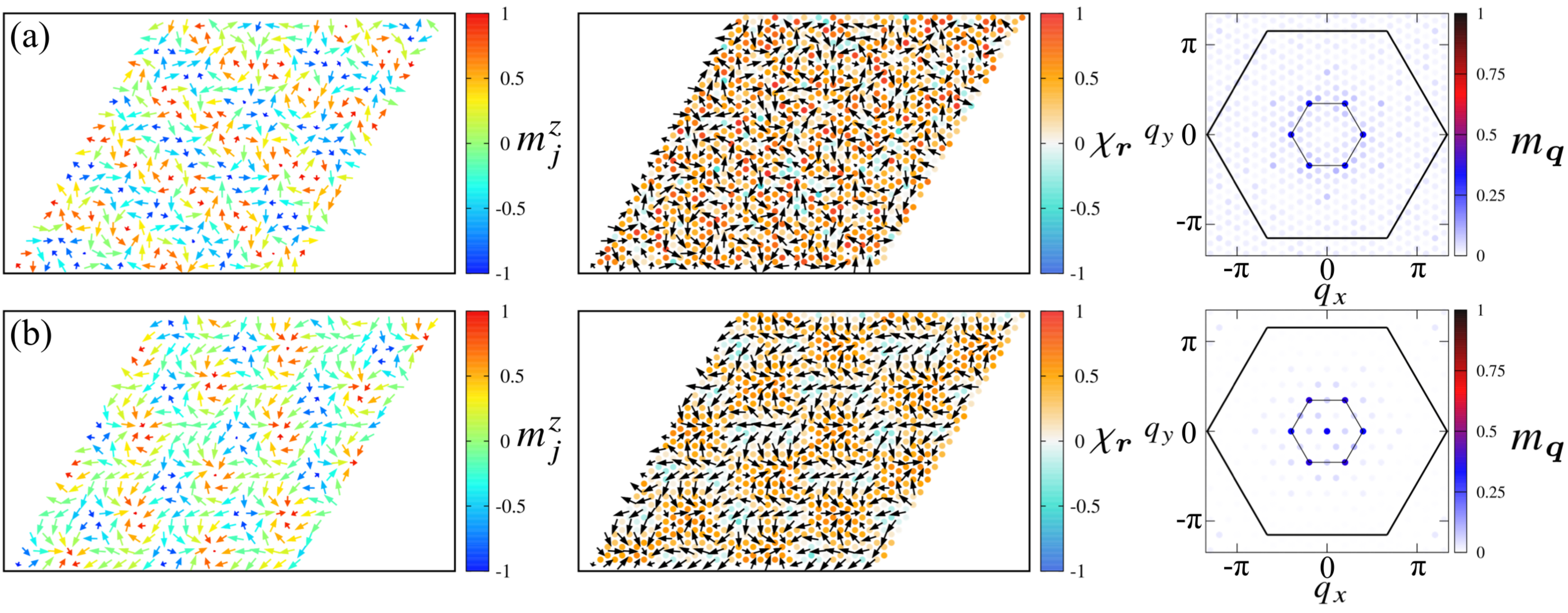} 
\caption{\label{fig:RCP}
Single snapshots of the NESSs by the right circularly polarized electric fields with (a) $\Omega/|E_{1Q}|=7$ and $E_d/|E_{1Q}|=1.0$ and (b) $\Omega/|E_{1Q}|=0.15$, $E_d/|E_{1Q}|=0.05$.
(Left panels) Magnetic configurations; the arrows and color denote the magnetic moments $\bm{m}_j(t)$ and the $z$ component $m^z_j(t)$, respectively.
(Middle panels) Scalar spin chirality configurations; the arrows and color of the circles denote the magnetic moments $\bm{m}_j(t)$ and the scalar spin chirality $\chi_{\bm{r}}(t)$, respectively.
(Right panels) Magnetic moments in momentum space $m_{\bm{q}}(t)$.
The large hexagons represent the first Brillouin zone.
The vertices of the small hexagons correspond to $\pm\boldsymbol{q}^*_1$, $\pm\boldsymbol{q}^*_2$, and $\pm\boldsymbol{q}^*_3$.
}
\end{center}
\end{figure*}

In Sec.~\ref{sec:simulation}, we show the NESSs by the electric field with the LCP.
We confirm that similar results are obtained by radiating the electric field with the RCP.
The difference between the NESSs by the electric field with the LCP and the RCP is the sign of the skyrmion number and the scalar spin chirality.
We show the snapshots of the NESS by the right circularly polarized electric field with $\Omega/|E_{1Q}|=7$ and $E_d/|E_{1Q}|=1.0$ in Fig.~\ref{fig:RCP}(a).
This state corresponds to the higher-order SkX with the positive skyrmion number $N_{\rm sk}=+24$; the sign of $N_{\rm sk}$ is accounted for by the negative three-spin interactions induced by the electric field with the RCP in Eq.~(\ref{eq:floquet_hamiltonian}).   
We also show the snapshots of the NESS by the right circularly polarized electric field with $\Omega/|E_{1Q}|=0.15$ and $E_d/|E_{1Q}|=0.05$ in Fig.~\ref{fig:RCP}(b), which corresponds to the bimeron crystal with $N_{\rm sk}=+12$.

\section{Derivation of the Floquet Hamiltonian}
\label{sec:derivation}

We derive the Floquet Hamiltonian in Eq.~(\ref{eq:floquet_hamiltonian}) by adopting the high-frequency expansion for classical spin systems under a time-periodic field \cite{higashikawa2018floquet}.
In the LLG equation in Eq.~(\ref{eq:LLG}), the time-dependent effective magnetic field for $\bm{m}_j$ is given by 
\begin{align}
\label{eq:Beff}
\bm{B}^{\rm eff}_j (t) = &-\frac{\partial \mathcal{H}_0}{\partial \bm{m}_{j}
}-\lambda\sum_{k_{\rm NN}}  \bm{m}_{k_{\rm NN}} \times \{\bm{E}(t) \times \bm{e}_{jk_{\rm NN}}\}, 
\end{align} 
where $k_{\rm NN}$ represents the nearest-neighbor sites of the site $j$.
Since this form is time-periodic with a period of $T=2\pi/\Omega$, we can adopt the high-frequency expansion in the Floquet formalism.
Then, a time-independent effective magnetic field up to $\Omega^{-1}$ is given by
\begin{align} 
\label{eq:Beff_Floquet}
\tilde{\bm{B}}^{\rm eff}_j &= \bm{B}^{\rm eff}_{j,0} + \sum_{n > 0}\frac{i\gamma [\bm{B}^{\rm eff}_{j,-n},\bm{B}^{\rm eff}_{j,+n}]}{n \Omega (1+\alpha_{\rm G}^2)}.
\end{align} 
Here, the time-independent effective magnetic field is represented by using the Fourier component of the time-periodic effective magnetic field $\bm{B}^{\rm eff}_{j,n} = T^{-1}\int^{T}_{0} dt\bm{B}^{\rm eff}_{j}(t) e^{in \Omega t}$; $n$ is an integer.
The Fourier components become nonzero for $n= 0,\pm1$, which are given by
\begin{align}
\label{eq:B0}
\bm{B}^{\rm eff}_{j,0} =&-\frac{\partial \mathcal{H}_0}{\partial \bm{m}_j},\\
\label{eq:B1}
\bm{B}^{\rm eff}_{j,\pm1} =& 
-\lambda\sum_{k_{\rm NN}}  \bm{m}_{k_{\rm NN}} \times (\tilde{\bm{E}}_{\pm 1} \times \bm{e}_{jk_{\rm NN}}),
\end{align}
where $\tilde{\bm{E}}_{\pm 1}=E_0(\delta,\pm i,0)/2$. 
The relation $[A , B]$ is defined as
\begin{align} 
\label{eq:commun}
&[\bm{B}^{\rm eff}_{j,-n},\bm{B}^{\rm eff}_{j,+n}] \nonumber \\
&= \bm{B}^{\rm eff}_{j,-n}\times\bm{B}^{\rm eff}_{j,+n} +\sum_{k\neq j} \left[ ( \bm{B}^{\rm eff}_{k,-n} \cdot \bm{L}_k ) \bm{B}^{\rm eff}_{j,+n} \right.\nonumber \\
&\left.- ( \bm{B}^{\rm eff}_{k,+n} \cdot \bm{L}_k ) \bm{B}^{\rm eff}_{j,-n} \right] + \mathcal{O}(\alpha_{\rm G}),
\end{align} 
with $L_k^\alpha = - \sum_{\beta,\eta} \epsilon_{\alpha\beta\eta} m_{k}^\beta (\partial / \partial m_{k}^\eta)$ $(\alpha,\beta,\eta = x,y,z)$. 
According to Eqs.~(\ref{eq:Beff_Floquet})-(\ref{eq:commun}), we obtain the time-independent effective magnetic field.
The Floquet Hamiltonian in Eq.~(\ref{eq:floquet_hamiltonian}) is obtained from a relation $\tilde{\bm{B}}^{\rm eff}_j=-\partial\mathcal{H}^{\rm F}/\partial \bm{m}_j$. 
It is noted that we ignore the contribution from $\mathcal{O}(\alpha_{\rm G})$ in Eq.~(\ref{eq:commun}) by assuming $\alpha_{\rm G}\ll 1$.

\bibliography{main.bbl}
\end{document}